\documentclass[preprint,aps,singlecolumn,superscriptaddress,preprintnumbers,nopacs,floatfix,amsmath,amssymb,overcite,citeauthorscript]{revtex4}

\usepackage[pdftex]{graphicx}
\usepackage{dcolumn}
\usepackage{bm}

\usepackage{amsmath}
\usepackage{graphicx}
\usepackage{amssymb}
\usepackage{mathrsfs}
\usepackage{bbm}
\usepackage{epsfig}
\usepackage{MnSymbol}

\newcommand{\la}[1]{\mbox{$
\lefteqn{ \mbox{\,\, \tiny #1}}$} \label{#1}}
\def\thebibliography#1{{\section*{References}}\list
 {$^\arabic{enumi}$}{\settowidth\labelwidth{[#1]}\leftmargin\labelwidth
 \advance\leftmargin\labelsep
 \usecounter{enumi}}
 \def\newblock{\hskip .11em plus .33em minus .07em}
 \sloppy\clubpenalty4000\widowpenalty4000
 \sfcode`\.=1000\relax}

\setcitestyle{super}

\reversemarginpar

\newcommand{\be}{\begin{eqnarray}}
\newcommand{\ee}{\end{eqnarray}}

\begin{document}





\pagenumbering{arabic}
\renewcommand{\baselinestretch}{2}

%
\title{ASPECTS OF STRUCTURAL LANDSCAPE   \\ OF HUMAN ISLET AMYLOID POLYPEPTIDE}

\author{Jianfeng He}
\email{hjf@bit.edu.cn}
\affiliation{School of Physics, Beijing Institute of Technology, Beijing 100081, P.R. China
}
\author{Jin Dai}
\email{daijing491@gmail.com}
\affiliation{School of Physics, Beijing Institute of Technology, Beijing 100081, P.R. China
}
\author{Jing Li}
\email{jinglichina@139.com}
\affiliation{Institute of Biopharmaceutical Research, 
Yangtze River Pharmaceutical Group \\ Beijing Haiyan Pharmaceutical Co., Ltd, Beijing 102206, China}
\author{Xubiao Peng} 
\email{xubiaopeng@gmail.com}
\affiliation{Department of Physics and Astronomy, Uppsala University,
P.O. Box 803, S-75108, Uppsala, Sweden}
\author{Antti J. Niemi}
\email{Antti.Niemi@physics.uu.se}
\homepage{http://www.folding-protein.org}
\affiliation{
Laboratoire de Mathematiques et Physique Theorique
CNRS UMR 6083, F\'ed\'eration Denis Poisson, Universit\'e de Tours,
Parc de Grandmont, F37200, Tours, France}
\affiliation{Department of Physics and Astronomy, Uppsala University,
P.O. Box 803, S-75108, Uppsala, Sweden}
\affiliation{School of Physics, Beijing Institute of Technology, Beijing 100081, P.R. China
}

\begin{abstract}
\noindent
The human islet amyloid polypeptide (hIAPP) co-operates with insulin to maintain
glycemic balance. It also constitutes the amyloid plaques that aggregate in
the pancreas of type-II diabetic patients.  
We have performed extensive {\it in silico} investigations to analyse the structural 
landscape of monomeric hIAPP, which is presumed to be intrinsically disordered.
For this we construct from first principles  a highly predictive energy function that 
describes a monomeric  hIAPP 
observed in a NMR experiment, as a local energy minimum. 
We subject our theoretical model of hIAPP  to repeated heating and cooling 
simulations, back and forth between a high temperature regime  where the 
conformation resembles a random walker and a low temperature limit where 
no  thermal motions prevail.   We find that the final low temperature conformations  
display a high level of degeneracy, in a manner which is fully in line with the presumed 
intrinsically disordered character 
of hIAPP.    In particular, we identify an isolated  family of $\alpha$-helical  conformations 
that {\it might} cause the  transition  to amyloidosis, by nucleation.
\end{abstract}

%

\maketitle

\section{introduction}
The human islet amyloid polypeptide (hIAPP),  also known as amylin, 
is a widely studied 37 amino acid polypeptide hormone. \cite{Westermark-2011,Pillay-2013,Cao-2013}
hIAPP is processed in pancreatic $\beta$-cells, 
by a protease cleavage in combination of post-translational modifications. 
Its secretion  responses to meals, and  the peptide co-operates with 
insulin to regulate blood glucose levels. But hIAPP can also aggregate  
into pancreatic amyloid deposits. The formation and 
buildup of amyloid fibrils correlates strongly with the depletion of islet $\beta$-cells.
The hIAPP amyloidosis is present in 
over 90 per cent of the type-II diabetic  patients 
\cite{Kahn-1999,Hoppener-2000,Chiti-2006,Haataja-2008} and the deposits are
considered the hallmark of the disease in progression.
Early studies  \cite{Lorenzo-1994,Fox-1993}  suggested that the fibrils themselves could be the toxic agents that 
cause cell death.  However,  recently it has been found  that the formation of amyloid plaques is  most likely a
sufficient and  not a necessary condition for the disruption of $\beta$-cells \cite{Brender-2012a,Nath-2011}. 
It appears that the cause for the  islet $\beta$-cell depletion is somewhere upstream from the 
formation and buildup of amyloid fibrils. The initial step seems to be an intracellular process that 
takes place in endoplasmic reticulum, golgi or secretory granules \cite{Westermark-2011,Pillay-2013,Cao-2013}.

Experimentally, the structure of hIAPP amyloid fibrils has been studied extensively.  See for example 
\cite{Kayed-1999,Harper-1997,Sumner-2004,Jayasinghe-2004,Jayasinghe-2005,Meetoo-2007,Sawaya-2007,Luca-2007,Nanga-2008,Wiltzius-2008,Wiltzius-2009,Dupuis-2009,Patil-2009,Nanga-2011,Nanga-2011a,Li-2012,Bedrood-2012,Salamekh-2012,Pannuzzo-2013,Brender-2013,Patel-2014}.
We note that the fibrils consist of an ordered parallel arrangement of hIAPP monomers, with
the cross-$\beta$ spine  displaying a
zipper-like packing.  Atomic level investigation
of  the cross-$\beta$ spine reveals 
that the segment which consists of residues 21-27 (NNFGAIL) forms a turn, that joins 
sheets which are made up of the residues 28-33 (SSTNVG), into a classic steric 
zipper \cite{Wiltzius-2008}. 
According to \cite{Li-2012,Bedrood-2012}  the fibril formation proceeds by nucleation,
so that one hIAPP molecule first assumes a hairpin  structure with  two $\beta$-strands linked by a loop.
This is  followed by a piling-up of  monomers. However, the structure of a full-length {\it monomeric}  hIAPP,
and in particular the intra-cellular conformational pathways that lead to
the $\beta$-hairpin nucleation causing conformation, remain
unknown \cite{Cao-2013}.  The  sole crystallographic structure  with 
Protein Data Bank (PDB) 
access code 3G7V \cite{Wiltzius-2009} describes
hIAPP fused with a maltose-binding protein. Two solution NMR structures are available in PDB.  
The PDB access codes are  2KB8  \cite{Patil-2009} 
and 2L86 \cite{Nanga-2011}. These three presently available PDB structures  are all very different 
from each other. Indeed, an isolated hIAPP is presumed to be an example of a dynamical, intrinsically disordered 
protein \cite{Dunker-2001}. When biologically active, such proteins are often presumed to be in a perpetual motion, 
fulfilling their biological function by constantly varying their shape. Thus these 
proteins lack an ordered folded conformation that could be studied {\it e.g.} by conventional x-ray 
crystallography approaches. Moreover,  detergents such as SDS micelles that are introduced 
as stabilising agents in solution NMR experiments, may  lead to structural distortions. 


The detailed atomic level structure of hIAPP conformations could in principle be extracted 
using molecular dynamics simulations.
Indeed, both all-atom and coarse-grained  molecular dynamics force fields 
are being employed to try and understand {\it in silico} the structure of hAIPP, both in fibrils and in isolation; see {\it e.g.}\cite{Li-2013,Xu-2012,Qiao-2013,Wu-2013,Miller-2013,Pannuzzo-2013,Liang-2013,Berhanu-2014}, and we refer to 
\cite{Berhanu-2014b} for  a recent detailed review.  However, in particular 
when explicit water is introduced in the simulations, the computational task becomes staggering: 
The special purpose molecular dynamics machine {\it Anton} \cite{Shaw-2008,Lindorff-Larsen-2012}  is capable of
describing {\it in vitro/in vivo} trajectories up to around a microsecond per a day {\it in silico}. 
At the same time, amyloid aggregation takes hours, even days. Thus the quality of present MD 
based investigations of hIAPP depends largely on our ability to determine the initial conformation 
in the simulations \cite{Berhanu-2014b}.

In the present article we investigate computationally the structural properties of the hIAPP segment, that consists
of  the  residues 9-37 where several studies have either observed 
or predicted that the amyloid fibril formation starts \cite{Westermark-2011,Pillay-2013,Cao-2013,Wang-1993,Westermark-1990,Nilsson-1999,Tenidis-2000,Goldsbury-2000,Azriel-2001,Jaikaran-2001,Scrocchi-2002,Mazor-2002,Scrocchi-2003,kajava-2005,Galzitskaya-2006,Luca-2007,Zhang-2007,Shim-2009}.
The physical properties of the 
short N-terminal segment that comprises the residues 1-8 is not addressed here. Its structure 
is more involved, due to the disulfide bond that forms between the cysteines which  are located at the residues 
2 and 7 \cite{Williamson-2009,Vaiana-2009}. Moreover, it remains to be understood what is the r\^ole
of the residues 1-8 in hIAPP aggregation \cite{Cope-2013}.
These residues appear to have a tendency towards
forming long and stable non-$\beta$-sheet fibers in solution, under the same conditions in which hIAPP 
aggregates into amyloid fibers. 
We note that a peptide, which consists only of the sites 17-37 of hIAPP, has also been identified both in 
human pancreas and plasma \cite{Nakazato-1990}. But its biological r\^ole  remains to be clarified.

Our approach is based on an universal energy function \cite{Niemi-2003,Danielsson-2010}; see \cite{Niemi-2014} for a detailed description. 
This approach builds on the powerful techniques of universality and renormalisation 
groups \cite{widom-1965,kadanoff-1966,wilson-1971,fisher-1974}  in combination with the notion of 
local gauge symmetry. Instead of a short time step expansion on which the MD approaches are based, 
we expand in terms of  variables that have slow spatial variations; in the continuum 
limit this becomes an expansion in terms of derivatives. As such our approach should provide complementary information
to MD approaches. In particular,  since the notion of a short time step 
is avoided we can in principle cover very long time scales.
 
 We note that the technique of universality was originally introduced to describe phase transitions and critical phenomena, while the method of renormalization group originates from high energy physics. Both have subsequently found numerous applications for example in dynamical systems and chaos, in statistical polymer research,  and 
in analysis of nonlinear ordinary and partial differential equations. 
%

We use the NMR structure 2L86 \cite{Nanga-2011} as a decoy to train the energy 
function.  The 2L86 is measured at pH of 7.4 {\it i.e.} around
the pH value in the extracellular domain,  where the actual amyloid deposit aggregation takes place.
We follow \cite{Chernodub-2010,Molkenthin-2011,Hu-2011,Krokhotin-2012a,Krokhotin-2012b,Krokhotin-2013a,Krokhotin-2013b,Hu-2013,Ioannidou-2014} 
to construct 
a static multi-kink configuration as an extremum of the energy function, so that it
accurately describes  the hIAPP structure in 2L86.  

The  2L86 is a composite of hIAPP with SDS micelles. It is often thought that SDS micelles could
model the  effects of a cell membrane. Thus our simulations correspond to
the  following biological set-up: 
We consider the structural evolution of an isolated  hIAPP  
in the extracellular domain  where it  has the initial shape of 2L86, and is 
in an initial interaction with the cell membrane. We  study the evolution of the hIAPP
conformation as it departs the cell membrane.
For this we inquire whether there are local energy minima, with a lower
energy  than that of the multi-kink which models 2L86. 

We subject the multi-kink to a series of  heating and cooling 
simulations \cite{Krokhotin-2012b,Krokhotin-2013a,Krokhotin-2013b}.
During the heating, we increase the temperature until we detect 
a structural change in the multi-kink, so that the configuration behaves like
a random walker, and we fully thermalise the configuration  
at the random walk temperature. The heating enables the multi-kink
to cross over the energy barriers which surround the initial 2L86 conformation,
in search of lower energy states.
We then   
reduce the ambient temperature, to cool down the configuration 
to {\it very} low temperature values until it
freezes into a conformation where no thermal motion prevails.
According to Anfinsen \cite{anfinsen-1973}, upon cooling
the protein should assume a fold, which is a local minimum 
of the low temperature thermodynamic (Helmholtz) free energy. More specifically,  
in the case of a protein with an ordered native fold,
the heating and cooling cycle should produce a highly localized statistical distribution of
structurally closely related conformational substates. When taken together, this ensemble 
constitutes the folded native state at low temperatures \cite{Frauenfelder-1988}.  In the case of  the energy function
introduced in \cite{Chernodub-2010,Molkenthin-2011,Hu-2011,Krokhotin-2012a} this has been shown to occur 
{\it in silico}, with a 
number of proteins that are known to possess an ordered native fold  \cite{Krokhotin-2012b,Krokhotin-2013a,Krokhotin-2013b}.

But for a protein which is  intrinsically disordered, instead we expect that the low temperature limit 
produces a {\it scattered} statistical distribution of structurally disparate but
energetically comparable ensembles of conformational substates.  Moreover, these different substates should be 
separated from each other by relatively low energy barriers.  
We propose that the  unstructured, disordered character of the protein is a consequence of a 
motion around this  landscape:  The protein swings and sways back and 
forth, quite freely,  over the low energy barriers that separate the various energetically 
degenerate but structurally disparate conformations.  We now proceed to show  that  in the case of hIAPP, 
heating and cooling procedure yields exactly this kind of 
structurally scattered ensembles of conformations.

\section{Methods}

\subsection{C$\alpha$ backbone}

Let $\mathbf r_i$ be the skeletal  C$\alpha$ coordinates of a protein, with $i=1,...,N$.
Introduce
the unit tangent vectors $\mathbf t_i$, unit binormal vectors $\mathbf b_i$,
and unit normal vectors $\mathbf n_i$
\begin{equation}
\mathbf t_i = \frac{ {\bf r}_{i+1} - {\bf r}_i  }{ |  {\bf r}_{i+1} - {\bf r}_i | } \ \ \ \ \ \& \ \ 
\ \ \ \mathbf b_i = \frac{ {\mathbf t}_{i-1} \times {\mathbf t}_i  }{  |  {\mathbf t}_{i-1} 
\times {\mathbf t}_i  | } \ \ \ \ \ \& \ \ \ \ \ \mathbf n_i = \mathbf b_i \times \mathbf t_i  
\label{tbn}
\end{equation}
The orthonormal triplet ($\mathbf n_i, \mathbf b_i , \mathbf t_i$) determines the discrete Frenet  
frame \cite{Hu-2011b} at the position $\mathbf r_i$ of the backbone; see figure \ref{fig-1}.
The C$\alpha$ backbone bond  and torsion angles, shown in Figure \ref{fig-2},  are computed as
follows
\marginpar{Fig \ref{fig-1}}
\marginpar{Fig \ref{fig-2}}
\begin{eqnarray}
\kappa_{i} \ \equiv \ \kappa_{i+1 , i} & = & \arccos \left( {\bf t}_{i+1} \cdot {\bf t}_i \right) \\
\tau_{i} \ \equiv \ \tau_{i+1,i} & = & {\rm sign}\{ \mathbf b_{i} \times \mathbf b_{i+1} \cdot \mathbf t_i \}
\cdot \arccos\left(  {\bf b}_{i+1} \cdot {\bf b}_i \right) 
\label{bondtors}
\end{eqnarray}
Alternatively, if these angles are all known, 
we can use  the discrete Frenet equation \cite{Hu-2011b}
\begin{equation}
\left( \begin{matrix} {\bf n} \\  {\bf b } \\ {\bf t} \end{matrix} \right)_{i+1}
= 
\left( \begin{matrix} \cos\kappa \cos \tau & \cos\kappa \sin\tau & -\sin\kappa \\
-\sin\tau & \cos\tau & 0 \\
\sin\kappa \cos\tau & \sin\kappa \sin\tau & \cos\kappa \end{matrix}\right)_{\hskip -0.1cm i+1 , i}
\left( \begin{matrix} {\bf n} \\  {\bf b } \\ {\bf t} \end{matrix} \right)_{i}  \label{DFE2}
\end{equation}
to construct all the frames along the entire C$\alpha$ backbone chain.  
Once we have constructed all the frames, we obtain the backbone coordinates from
\begin{equation}
\mathbf r_k = \sum_{i=0}^{k-1} |\mathbf r_{i+1} - \mathbf r_i | \cdot \mathbf t_i
\label{dffe}
\end{equation}
In  (\ref{dffe}) we may set $\mathbf r_0 = 0$, and we may choose $\mathbf t_0$ to point 
along the positive $z$-axis. With the exception of $cis$-proline which is rare, we
may take the distance between any two neighboring  C$\alpha$ atoms to have a constant value
\[
 |\mathbf r_{i+1} - \mathbf r_i | \ \approx \ 3.8 \ {\rm \AA}
 \] 
This approximation is valid at time scales which are much longer than the characteristic time scale of covalent bond vibration;
here we are interested in the limit of such long time scales. 
For any
two C$\alpha$ atoms that are not next to each other along the backbone chain, the PDB structures are
consistent with  the steric constraint
\begin{equation}
|\mathbf r_i - \mathbf r_k | > 3.8 \ {\rm \AA} \ \ \ \ \ \ {\rm for} \ \ \ | i - k | \geq 2
\la{steric}
\end{equation}
The positions of the backbone N, C, O and H atoms and  the side-chain C$\beta$ atoms
can be determined quite precisely in terms of the C$\alpha$ coordinates\cite{Schrauber-1993,Dunbrack-1993,Lovell-2000,Dunbrack-2002,Shapovalov-2011,Lundgren-2012c,Lundgren-2012d}. 
Similarly, several higher level side-chain atoms assume rotamer positions with respect to the C$\alpha$ backbone.
\cite{Schrauber-1993,Dunbrack-1993,Lovell-2000,Dunbrack-2002,Shapovalov-2011,Lundgren-2012c,Lundgren-2012d}. 
The C$\alpha$ backbone  is also widely 
exploited in structural classification schemes such as SCOP \cite{scop} and CATH \cite{cath}, in 
homology modeling \cite{Zhang-2009}, in {\it de novo} approaches \cite{Dill-2007}, 
and in the development of coarse grained energy functions for folding 
prediction \cite{Scheraga-2007}.  
The goal of the so called C$\alpha$-trace 
problem  \cite{DePristo-2003,Lovell-2003,Rotkiewicz-2008,
Li-2009,Krivov-2009} is to construct an accurate 
all-atom model of the natively folded protein from 
the positions of the central C$\alpha$ atoms. 
Both knowledge-based approaches such as MAXSPROUT \cite{DePristo-2003}  and {\it de novo} 
methods including PULCHRA \cite{Rotkiewicz-2008}  and  SCWRL \cite{Krivov-2009}
address  the C$\alpha$-trace problem. It has been shown that the virtual  C$\alpha$ backbone  
bond  ($\kappa_i$) and torsion ($\tau_i$)  angles 
are sufficient to determine the structure of any protein in Protein Data Bank, with very high precision 
\cite{Hinsen-2013}.  Thus, the C$\alpha$ coordinates form an attractive set of variables to try and
approximate the structure and dynamics of proteins.

\subsection{Universal energy function} 

A folded protein minimises locally the  thermodynamical Helmholtz free energy  
\begin{equation}
F = U - T S
\la{Helm}
\end{equation}
where $U$ is the internal energy, $S$ is the entropy, and
$T$ is the  temperature.  The free energy is a  function of all the  inter-atomic distances  
\begin{equation}
F = F(r_{\alpha\beta})\ ; \ \ \ r_{\alpha\beta} = |\mathbf r_\alpha - \mathbf r_\beta |
\la{intat}
\end{equation}
where index $\alpha,\beta, ... $ extends over all the atoms in the protein system.
In the case of slowly varying  deformations,  we may 
follow the general  {\it universality} arguments in \cite{widom-1965,kadanoff-1966,wilson-1971,fisher-1974}.  
These arguments instruct us to adopt the C$\alpha$ backbone bond and torsion angles as the structural order 
parameters to characterise the protein conformation  in the vicinity of the free energy 
minimum,
\[
r_{\alpha\beta} \ = \  r_{\alpha\beta} (\kappa_i, \tau_i)
\]
Accordingly, when deformations around a minimum energy configuration remain  slow and small
we may expand the free energy in 
terms of the C$\alpha$ bond and torsion angles. In 
\cite{Niemi-2003,Danielsson-2010,Chernodub-2010,Molkenthin-2011,Hu-2011,Krokhotin-2012a,Krokhotin-2012b,Krokhotin-2013a,Krokhotin-2013b}
it has been shown that in the  limit of  slowly varying variables,
 the free energy $F$ expands as follows:
\begin{equation}
F  = - \sum\limits_{i=1}^{N-1}  2\, \kappa_{i+1} \kappa_{i}  + 
+ \sum\limits_{i=1}^N
\biggl\{  2 \kappa_{i}^2 + \lambda\, (\kappa_{i}^2 - m^2)^2  
\
\  
+ \frac{q}{2} \, \kappa_{i}^2 \tau_{i}^2   
- p \,  \tau_{i}   +  \frac{r}{2}  \tau^2_{i} 
\biggr\} \ + \dots
\label{E1old}
\end{equation}
Here $\lambda$, $q$, $p$, $r$, and $m$ 
depend on the atomic level physical properties and the chemical 
microstructure of the protein and its environment, and in principle these parameters can 
be computed from this knowledge.
 We note the following:  The free energy (\ref{E1old}) is a variant of the energy function of the discrete 
nonlinear Schr\"odinger  equation (DNLS)  \cite{Faddeev-1987,Ablowitz-2004}: The first sum 
together with the three 
first terms in the second sum is the energy of  the standard  DNLS equation,  
in terms of the discretized Hasimoto variable \cite{Hasimoto-1972}. 
 The fourth term ($p$) is the conserved "helicity", it is responsible for the chirality of the C$\alpha$ backbone.
The last ($r$) term is the Proca mass.  

 {\it  A priori}, the fundamental range of the bond angle $\kappa_i$ is  $ [0,\pi]$. For the 
torsion angle the fundamental range is  $\tau_i \in [-\pi, \pi)$. Consequently ($\kappa_i, \tau_i$) can be identified
with the canonical 
latitude and longitude angles on the surface of a sphere. 
In the sequel we find it useful to extend the fundamental range
of $\kappa_i$ into $ [-\pi,\pi]$  but with no change in the fundamental range of $\tau_i$. 
We compensate for this two-fold covering of the sphere, 
by the following discrete $\mathbb Z_2$ symmetry \cite{Hu-2011b}
\begin{equation}
\begin{matrix}
\ \ \ \ \ \ \ \ \ \kappa_{l} & \to  &  - \ \kappa_{l} \ \ \ \hskip 1.0cm  {\rm for \ \ all} \ \  l \geq i \\
\ \ \ \ \ \ \ \ \ \tau_{i }  & \to &  \hskip -2.5cm \tau_{i} - \pi 
\end{matrix}
\la{dsgau}
\end{equation}
Finally, we note that regular protein secondary 
structures  correspond to constant values of
$(\kappa_i, \tau_i)$.  For example standard $\alpha$-helix and $\beta$-strand are
\begin{equation}
\alpha-{\rm helix:} \ \ \ \ \left\{ \begin{matrix} \kappa \approx \frac{\pi}{2}  \\ \tau \approx 1\end{matrix} \right.
\ \ \ \ \ \ \& \ \ \ \ \ \ 
\beta-{\rm strand:} \ \ \ \ \left\{ \begin{matrix} \kappa \approx 1 \\ \tau \approx \pi \end{matrix}  \right.
\la{bc2}
\end{equation}
Similarly,  all the other regular secondary structures such as 3/10 helices, 
left-handed helices {\it etc.} are structures with definite constant values of $\kappa_i$ and $\tau_i$.
A loop can be defined to be 
any ($\kappa_i, \tau_i$) configuration that interpolates between the regular structures. 
Along a loop the values of ($\kappa_i, \tau_i$) are variable, from site to site.

%
%
%
%
%
%
%
%
%
%
%
%
%
%
%

\subsection{Training the energy}

All the parameters in (\ref{E1old}) are in principle computable
from the atomic level knowledge of the protein, the solvent, and
the environmental characteristics including temperature, pressure, acidity and so forth. These 
parameters can also be estimated from molecular dynamics simulations, or by comparison 
with experimentally known structures. Here we use the experimental structure: 
We consider the residues 9-37 in the hIAPP  polypeptide with Protein Data Bank (PDB) code 2L86. We train the
energy function (\ref{E1old}) to model this configuration.

The PDB entry 2L86 consists of 20 NMR different configurations. 
In Figure \ref{fig-3}
\marginpar{Fig \ref{fig-3}}
we show an interlaced summary of these configurations.
Note the presence of  substantial fluctuations in the residues 1-8 of the N-terminal.
The residues 9-37 form a much more stable structure; as can be seen in Figure 3 the 
variations between the NMR structures are minor, over the sites 9-37. 
Any of the 20 different NMR structures could be utilized,
to train the energy function (\ref{E1old}). A structure obtained {\it e.g.} by averaging the NMR structures 
could also be used. The differences between
these choices are minor, and  for concreteness we use here 
the first NMR structure in the PDB entry 2L86.
We determine the parameters so 
that the extremum of (\ref{E1old}) coincides with the profile of the 2L86:
A variation of (\ref{E1old})  with respect to $\tau_i$ gives
\[
\frac{\partial F}{\partial\tau_i} \ = \ d \kappa_i^2 \tau_i + c  \tau_i - a - 
b\kappa^2_i \  = 0 
\]
\begin{equation}
\Rightarrow \ \  \tau_i [\kappa] =   \frac{ a + b\kappa_i^2}{c + d\kappa_i^2}
\label{Etau}
\end{equation}
We evaluate the derivative of the energy  with respect to $\kappa_i$.  We  
substitute $\tau_i[\kappa]$ from  (\ref{Etau}) into the ensuing 
equation. We arrive at  the following  modified version of the  DNLS equation
\begin{equation}
\kappa_{i+1} - 2 \kappa_i + \kappa_{i-1} \ = \ U' [\kappa_i] \kappa_i 
\ \equiv\ \frac{dU[\kappa]}{d\kappa_i^2} \ \kappa_i \ \ \ \ (i=1,...,N)
\label{Ekappa}
\end{equation}
(with $\kappa_{0} = \kappa_{N+1} = 0$)
where
\[
U[\kappa] = -  
\frac{1}{d} \left( a d - bc \right)^2   \frac{1}{c + d\kappa^2}  \ - \ 
\frac{1}{2d} \left( b^2 + 4 q d m^2 \right) \, \kappa^2 + 
q\, \kappa^4
\]
This equation coincides with  the stationary points of the energy function 
\[
H \ = \  - 2\sum\limits_{i=1}^{N-1} \kappa_{i+1} \kappa_i \ + \ \sum\limits_{i=1}^N  \left\{ 2 \kappa_i^2 + U[\kappa_i] \right\}
\]
For proper parameter values the equation (\ref{Ekappa}) supports a kink solution.  The explicit form of the kink, in terms of elementary 
functions, is not known. But an approximation  can be  constructed 
numerically {\it e.g.} by using the iterative procedure described in reference 
\citenum{Molkenthin-2011}. We use the program package {\it ProPro} which is based on this iterative procedure, and
described in
\begin{equation}
{\rm http://www.folding-protein.org}
\label{propro}
\end{equation}
to construct the parameters. 

%
%
%
%
%
%
%
%
%
%
%
%
%
%
%

\subsection{Heating and cooling}
We study the energy landscape of the hIAPP by subjecting the energy function  (\ref{E1old}) that we have trained with the NMR structure 2L86 of hIAPP, 
to extensive heating and cooling simulations. It has been argued that in the case of simple proteins, the folding dynamics 
follows Arrhenius law. On the other hand, a simple spin system with dynamics determined by Glauber protocol 
\cite{Glauber-1963,Bortz-1975,Martinelli-1994a,Martinelli-1994b}
is also subject to Arrhenius law. This proposes that we  try and describe out-of-thermal-equilibrium 
dynamics of hIAPP using a combination of (\ref{E1old}) with Glauber protocol. The transition 
probability $\mathcal P(i \to j)$ between any two states $i$ and $j$ is evaluated from 
\begin{equation}
\mathcal P = \frac{x}{1+x} \ \  \ \ {\rm with}  \   \ \ \ x =     \exp\{ - \frac{ \Delta E}{kT} \}  
\label{P}
\end{equation}
The energy difference $\Delta E$ between the two states is computed 
from the free energy (\ref{E1old}).  We take all  parameters in (\ref{E1old}) to be temperature independent. 
As a consequence the temperature factor  $kT$ is not directly related to the physical temperature factor $k_B t$. But it
can be related to the physical temperature by 
the renormalization procedure detailed in \cite{Krokhotin-2013a}. Here
we are interested in  the low temperature limit energy landscape, thus 
the renormalization has no practical significance.

A full heating and cooling cycle involves $5 \times 10^7$ Monte Carlo steps,
which we have found to be adequate.
The cycle starts with a very low temperature value. After a preliminary
thermalization of the initial configuration during $5\times 10^6$ steps 
in the low temperature regime, we proceed to increase the temperature during $10^7$
MC steps. This is followed by a high temperature thermalization during $2 \times 10^7$ 
MC steps, after which we cool the system down during $10^7$ MC steps to the initial low 
temperature value where we then allow it to become fully thermalized. 
We have tested a number of different alternatives but {\it e.g.} an increase in the number of steps does not have any observable
effect on the results.  

Glauber dynamics is known to provide a quasi-realistic temporal evolution of a non-equilibrium process where
the heating and cooling proceeds slowly, with respect to the atomic scale.
Thus we trust that in combination with our universal
energy function, the evolution we obtain with  (\ref{P}) describes the universal statistical aspects  
of hIAPP  trajectories over biologically relevant distance and time  scales. 

 \section{Results}

\subsection{The three-kink solution} 

In Table I we list the parameter values,  that find by  training the energy function (\ref{E1old}) 
to describe 2L86. Note that there are only 21 parameters, while there are a total of 28 amino acids in 2L86
{\it i.e.} there are {\it less} parameters than there are amino-acids !
This implies that  the physical principles from  which the energy function (\ref{E1old}) derives, 
can be subjected to  {\it very} stringent experimental scrutiny.
\marginpar{Table I}
Note also that  in those terms of (\ref{E1old}) that engage the 
torsion angles, the numerical parameter values are consistently 
much smaller than in those  terms that involve only the bond angles. 
This is in line with the observation, that in proteins the 
torsion angles  {\it i.e.} dihedrals 
are usually quite flexible  while the bond angles are relatively stiff. 

In Figure \ref{fig-4} a)
\marginpar{Fig \ref{fig-4}}
we show the spectrum of bond and torsion angles for the first  NMR structure of 
2L86, with the convention that the bond angle takes values between $\kappa \in [0,\pi]$. 
In Figure \ref{fig-4} b) we have introduced the $\mathbb Z_2$ symmetry (\ref{dsgau}) to disclose that there are
three individual kinks along the backbone. The first kink from the N-terminal
is centered at the site 17. The third kink is centered at
the site 27. Both of these two kinks correspond to clearly visible  loops  in the three dimensional 
structure,  seen in Figure \ref{fig-3}.
The second kink, centered at site 23, is much less palpable in the 
three dimensional NMR structure. This kink  appears  more like a bend in an $\alpha$-helical 
structure,  extending from the first kink to the third kink.
This $\mathbb Z_2$ transformed ($\kappa,\tau$) profile in Figure \ref{fig-4} b) 
is the background in  (\ref{E1old}).

In Figure \ref{fig-5} we compare the bond and torsion angle spectrum of our three-kink solution with 
the first NMR structure of 2L86;  the solution  is  obtained by numerically 
solving the equations (\ref{Ekappa}), (\ref{Etau}) using the program {\it ProPro} 
described at (\ref{propro}).  
\marginpar{Fig \ref{fig-5}}
Clearly, the quality of our three-kink solution is  very good, at the level of the bond and torsion angles.

Figure \ref{fig-6} shows our three-kink solution, interlaced with the first NMR structure of 2L86.
\marginpar{Fig \ref{fig-6}}
The RMSD distance between the experimental 
structure and the three-kink configuration is 1.17 \AA. This is somewhat large, when compared
to the multi-kink structures that have been presented
in \cite{Krokhotin-2012b,Krokhotin-2013a,Krokhotin-2013b}; typically a multi-kink configuration 
describes a high resolution crystallographic structure with a  RMSD precision much below 1.0 \AA.
But the resolution of the present experimental NMR structure  is not that good,  
and this is reflected by the somewhat lower
quality of the three-kink solution, in comparison to the case of high resolution crystallographic structures.

Figure \ref{fig-7} 
\marginpar{Fig \ref{fig-7}}
compares the residue-wise C$\alpha$ distances, between the 20 different NMR structures in the
PDB entry 2L86, and our three-kink solution. For those residues that precede 
the bend-like second kink which is  centered at site 23, 
the distance between the experimental structures and the numerically constructed
solution is relatively small.  There is a quantitative change
in the precision of the three-kink solution, that takes place after site 23:
The distance between the experimental structures 
and the three-kink solution increases after this residue. 
We propose that this change is due to the SDS micelles,  used in the experimental
set-up to stabilize hIAPP/2L86: Sodium dodecyl sulfate (SDS)  is widely used as a 
detergent, to enable NMR structure determination in the case
of proteins with high hydrophobicity \cite{Bell-1990,Seddon-2004,Prive-2007,Michaux-2008}.  The mechanism 
of SDS-protein interaction is not yet fully understood. But it is known that the hydrophobic 
tails of SDS molecules interact in particular with the hydrophobic core of a protein. 
These interactions are known to disrupt the native structure to the effect, that
the protein displays an increase in its $\alpha$-helical posture.These additional 
$\alpha$-helical structures tend to be surrounded by SDS micelles.

The residue site 23 of hIAPP is the highly hydrophobic phenylalanine. It is  followed by the 
very flexible glycine at site 24. Thus,  the apparently abnormal
bend which is located  at the site 23 and affects the quality of our three-kink configuration, could be
due to an interaction between the phenylalanine and the surrounding 
SDS micelles. We note  that a high sensitivity of the hIAPP conformation  to the phenylalanine at 
site 23 has been recorded in several studies \cite{Porat-2003,Marshall-2011,Profit-2013}.

An analysis of 2L86 structure using MOLPROBITY \cite{Chen-2010} suggests  
a propensity towards poor rotamers between the sites 23-36, {\it i.e.}
the region where the quality of our three-kink solution decreases.

A comparison with the statistically determined 
radius of gyration relation \cite{Krokhotin-2012c}
\begin{equation}
R_g \ \approx \ 2.29 \cdot N^{0.37}
\label{Rg}
\end{equation}
where $N$ is the number of residues, reveals that for 2L86 the value of $R_g \approx 9.2$ (over residues 
$N=9,...,36$) is somewhat high.
According to (\ref{Rg}), we expect a value close to $R_g \approx 7.9$ (with $N=28$ residues): The structure of
2L86 should be more compact.

We conclude that most likely the SDS-hIAPP interaction has deformed a loop which, in the absence of micelles,
should be located in the vicinity of the residue number 23. Probably,  the interaction with micelles has converted
this loop into a structure resembling a bend in an $\alpha$-helix. This interaction between hIAPP and SDS
interferes with our construction of the three-kink configuration, adversely affecting its precision.

\subsection{Heating and cooling of  hIAPP}

The Figures \ref{fig-8}
\marginpar{Fig \ref{fig-8}}
describe the evolution of the three-kink configuration  during repeated heating and cooling. 
The Figure \ref{fig-8} (top) shows the evolution of the radius of 
gyration, and the Figure \ref{fig-8} (bottom) shows the RMSD distance 
to the PDB structure 2L86. Both the average value and the one standard deviation from the average value, are shown.
During the cooling period we observe only one  transition, in both the radius of gyration and the RMSD. 
Thus, we are confident that at high temperatures we are in 
the random walk regime \cite{Krokhotin-2012b,Krokhotin-2013a,Krokhotin-2013b}. 
The profile of each curve in Figure \ref{fig-8} shows that the structures are
fully thermalized, both in the high temperature and in the low temperature regimes. 

We observe that the average final value of the radius of gyration 
$R_g \approx 7.8$ is an excellent match
with the prediction  obtained from  (\ref{Rg}). 
The final configurations are quite different from the initial configuration:
The RMSD distance between
the initial configuration and the average final configuration is around  $4.8$ \AA.
  
Figure \ref{fig-9} shows results for a representative simulation with $\sim$1.500 complete heating and cooling
cycles; an increase in the number of cycles does not  have a qualitative effect. 
The Figure shows the distribution of the final  conformations,  grouped according to their  
radius of gyration versus  end-to-end distance. The final conformations form clusters, and we have identified the six 
major clusters that we observe in our simulations. 
By construction, the clusters correspond to 
local extrema of the energy function we have constructed to model 2L86.
Five of the clusters, denoted 2-6 in the Figure, have an apparent spread, there is  a flat direction in the energy 
around its extremum.  The clusters 3 and 5 are also somewhat more scattered than the clusters 2, 4 and 6. 
Finally,  the cluster number 1 is a localized one.  Note that the initial conformation, marked with 
red triangle in the Figure \ref{fig-9}, does not appear among the final configurations.  It is apparently 
an unstable extremum of the energy.
\marginpar{Fig \ref{fig-9}}

In Figure \ref{fig-10} we display the average conformations in each of the six clusters, interlaced with each other and
the initial 2L86 configuration. In this Figure, the first two C$\alpha$ atoms from the N-terminus are made to coincide. 
We have maximized the alignment of the subsequent C$\alpha$ atoms, to the extent it is possible.  
The Figure reveals the presence of substantial conformational difference between 
the clusters. 
\marginpar{Fig \ref{fig-10}}

In Figures \ref{fig-11}
\marginpar{Fig \ref{fig-11}}
we compare the individual clusters shown in Figure \ref{fig-10}, with the initial 2L86 configuration (in blue). 
In each of these Figures,
we show ten representative entries in each of the clusters (in red), to visualize the extent of conformational fluctuations
within each cluster. We observe that the conformational spread within  each of the six clusters is not very large.

\subsection{Side-chain atoms}

We have reconstructed the full atom configurations for every structure shown in Figures \ref{fig-10}, \ref{fig-11}
using both 
PULCHRA \cite{Rotkiewicz-2008}  and  SCWRL \cite{Krivov-2009}.  We have excluded {\it all} 
configurations where {\it any} pair of heavy atoms either in the  backbone or in the side-chain,
that are {\it not} covalently bonded to each other,  have a mutual distance less than 1.2 \AA. In this manner we have
obtained the distributions shown in Figures \ref{fig-12}, for PULCHRA  and SCWRL respectively.
\marginpar{Fig \ref{fig-12}}
We  observe that qualitatively, the differences between Figures \ref{fig-9} and \ref{fig-12} are minor; the majority of the structures
that appear in Figure \ref{fig-9} are also consistent with the side-chain assignments given by  
both PULCHRA  and SCWRL, when combined with the requirement that any 
pair of atoms must have a minimum distance of 1.2 \AA~ unless they are covalently bonded; for the C$\alpha$ atoms
we have the bound (\ref{steric}) during our entire simulations.

Finally, in Figures \ref{fig-13} and \ref{fig-14} we display the hydrophobic side-chains, for the average configurations in
each of the six clusters shown in Figure \ref{fig-9}. 
\marginpar{Fig \ref{fig-13}}
\marginpar{Fig \ref{fig-14}}
We observe that the hydrophobic side-chains are by and large exposed to the solvent. A notable exception
is the pair L-16  and V-32 in cluster 1 which are {\it very} proximal   to each other. Note that I-26 is also close-by.
Since V and L are known to be mutually attractive \cite{Makowski-2008}, we 
conclude that the proximity of this pair enhances the stability of  the cluster number one.  

\section{Conclusions}

We have found that the three-kink configuration which models the C$\alpha$ backbone of the human islet amyloid polypeptide, 
is quite unsettled:  Its low temperature limit comes endowed with six different conformational clusters.  This 
is a marked contrast with the properties of a multi-kink configuration which models a protein that is 
known to possess a unique folded native state  \cite{Krokhotin-2012b,Krokhotin-2013a,Krokhotin-2013b}. 
But the low temperature clustering is in accord with the intrinsically disordered character of 
hIAPP: The different clusters can be viewed as instantaneous snapshot conformations, between which the dynamic 
hIAPP swings and sways in an apparently unsettled manner which is characteristic to any intrinsically disordered protein.  
Furthermore, an inspection of the side-chain atoms reveals that in five of the six clusters, 
the hydrophobic side-chains become quite exposed to the solvent. Thus protein-solvent interactions can be expected to be present, 
in a manner that further enhances the dynamic character of these clusters. 

But the cluster number one appears different.  
In this cluster the hydrophobic side-chains are less exposed.  In particular the L-16 and V-32 
in this cluster are positioned in a manner where they can be expected to interact attractively, 
in a manner that stabilizes the conformation.
Indeed, this cluster has a much more localized conformational distribution than the other five clusters, and the posture  comprises
of two anti-parallel helices. The cluster number one is a good candidate to trigger the formation
of hIAPP fibrils and amyloidosis. We propose that this cluster correspond to the intermediate $\alpha$-helical structures
observed {\it e.g.} in  \cite{Salamekh-2012,Cao-2013,Kayed-1999,Jayasinghe-2005,Harper-1997,Pannuzzo-2013}

\section*{Acknowledgements}
We thank A. Sieradzan for discussions. A.J.N. thanks A. Ramamoorthy for a communication and P. Westermark for numerous
discussions on hIAPP. We acknowledge support  from Region Centre 
Rech\-erche d$^{\prime}$Initiative Academique grant, Sino-French 
Cai Yuanpei Exchange Program (Partenariat Hubert Curien), 
Vetenskapsr\aa det, Carl Trygger's Stiftelse f\"or vetenskaplig forskning, and  Qian Ren Grant at BIT.

\newpage
\section*{Figure legends.}
\newcounter{fig}
%
%
\begin{list}
{\bf Fig. \arabic{fig}.}{\usecounter{fig}}
\item
Definition of Frenet frame. The tangent vector $\mathbf t_i$ points from a given C$\alpha$ atom
(labeled $i$) to the next C$\alpha$ atom (labeled $i+1$).
The normal vector $\bf n_i$ is perpendicular to $\bf t_i$ and lies in the plane defined by 
the ($i-1$)$^{th}$, $i$$^{th}$ and ($i+1$)$^{th}$ C$\alpha$ atoms. The bi-normal vector 
$\bf b_i$ completes the right-handed frame. 
\label{fig-1}

\item
Definitions of the backbone bond angle $\kappa_{i}$ and torsion angle $\tau_{i}$ 
in terms of the C$\alpha$ atoms. Note the indexing.
\label{fig-2}
\item
Interlaced NMR backbone structures of 2L86, from N-terminal (left) to  C-terminal (right).
\label{fig-3}
\item
a) The spectrum of the bond and torsion angles of 2L86 (first entry) with the convention that bond angle takes
values in $\kappa \in [0,\pi)$. b) The spectrum of the bond and torsion angles that identifies the
kink structures. 
\label{fig-4}
\item
Top: Comparison of the 3-kink bond angle (blue) with the experimental 2L86 bond angle spectrum (red).
Bottom: Comparison of the 3-kink torsion angle (blue) with the experimental 2L86 torsion angle spectrum (red).
\label{fig-5}
\item
The  3-kink solution (blue) interlaced with the 2L86 experimental structure (red).
\label{fig-6}
\item
The black line denotes the C$\alpha$ atom distance between the 3-kink configuration and the model 1 NMR configuration 2L86;
the grey region is  an estimated 0.15 \AA~ zero-point fluctuation distance from the 3-kink configuration. 
The red line denotes the B-factor Debye-Waller fluctuation distance from model 1 of 2L86. 
The blue-colored  points denote the average C$\alpha$ distance between the model 1 NMR structure from the average of the remaining
19 models on 2L86; the error-bars denote the maximal and minimal C$\alpha$ distances. 
\label{fig-7}
\item
The top figure shows how the radius of gyration of the three-kink configuration evolves during heating and cooling cycle. 
The bottom figure shows the same for the RMSD distance from the initial configuration (the PDB structure 2L86).
The red line is the average value over all configurations, and the grey zone marks the extent of the one standard 
deviation  from the average value. The Monte Carlo steps are displayed in multiplets of $10^6$.
\label{fig-8}
\item
The distribution of all final configurations, in a run with $\sim$1.500 full heating and cooling cycles, classified  
in terms of the radius of gyration and end-to-end distance of the final configuration. Each blue dot represents a single
final configuration.  The six major clusters are encircled with a black ellipse; a wider grey ellipse around the clusters 3 and 5
includes some nearby scattered states. The red triangle identifies the initial configuration, the entry 1 in 2L86. Note also the
presence of a cluster encircled with yellow between clusters 1 and 2.
\label{fig-9}
\item
Superposition of all the six major clusters in Figure \ref{fig-9},  interlaced with each other and with the PDB entry 2L86.
\label{fig-10}
\item
Superposition of ten representative conformations (red) in each of the six clusters, as marked, together with the PDB entry 2L86 (blue).
\label{fig-11}
\item
Conformations in Figure \ref{fig-9} that can be allocated side-chains with PUCLHRA (top) and SCWRL4 (bottom), with the additional
condition that the distance between any two heavy atoms that are not covalently bonded to each other exceed 1.2 \AA.  Note that the
entire cluster encircled with yellow and located between clusters 1 and 2 in Figure \ref{fig-9} is present in PUCLHRA but
excluded in SCWRL4.
\label{fig-12}
\item
Hydrophobic side chains for the clusters 1-3, assigned by PULCHRA. L light blue; A dark blue; F yellow; V green; I red.
In the cluster 1 the L-16 and V-32   are located very close to each other. All the other hydrophobic side-chains are exposed to
the solvent.
\label{fig-13}

\item
Hydrophobic side chains for the clusters 4-6, assigned by PULCHRA. L light blue; A dark blue; F yellow; V green; I red.
All the hydrophobic side-chains are exposed to the solvent.
\label{fig-14}

\end{list}
\clearpage
\resizebox{15.cm}{!}{\includegraphics[]{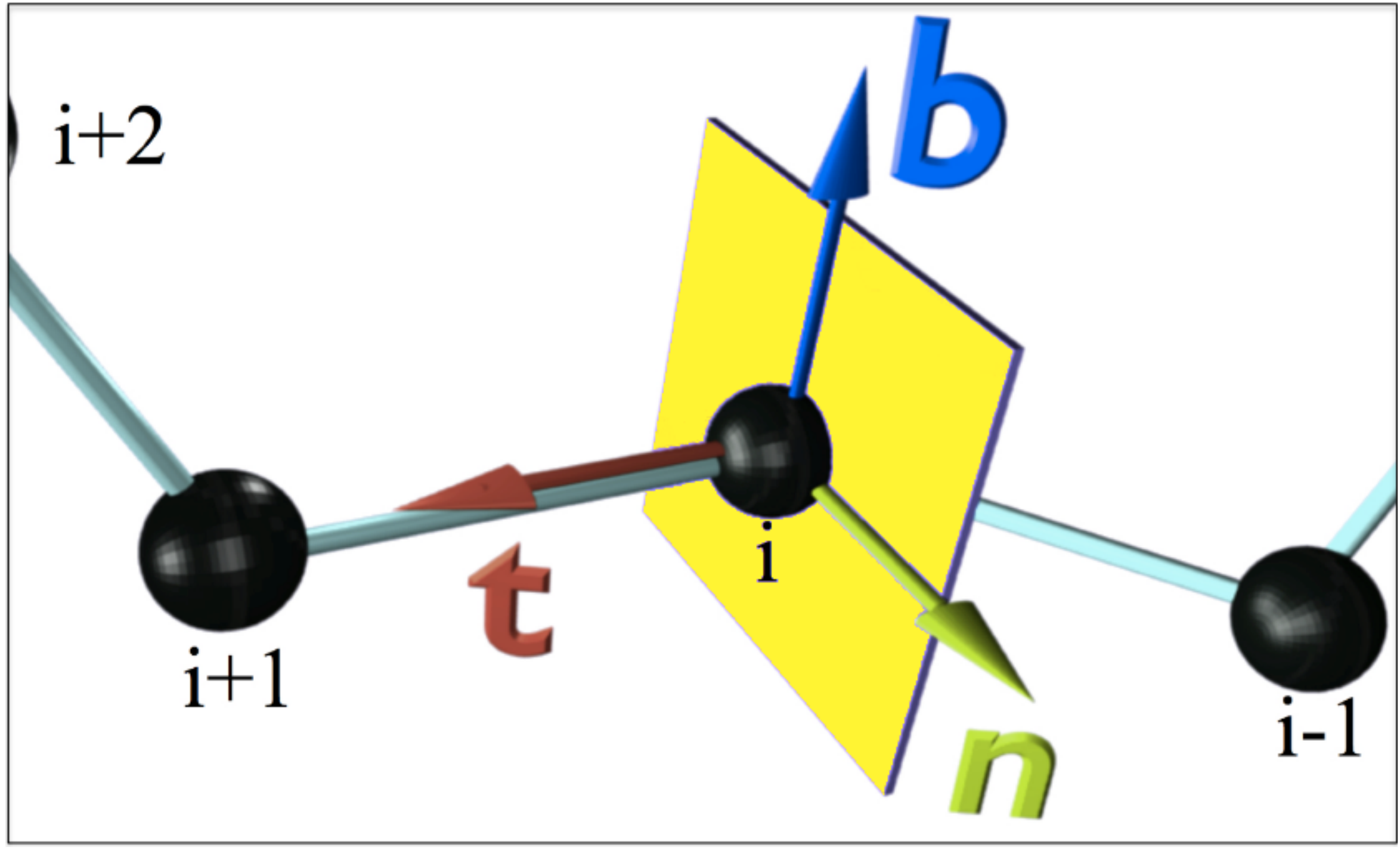}}
\vfill
\hfill{\large\sf Figure \ref{fig-1}}
\clearpage
\begin{center}
\resizebox{15.cm}{!}{\includegraphics[]{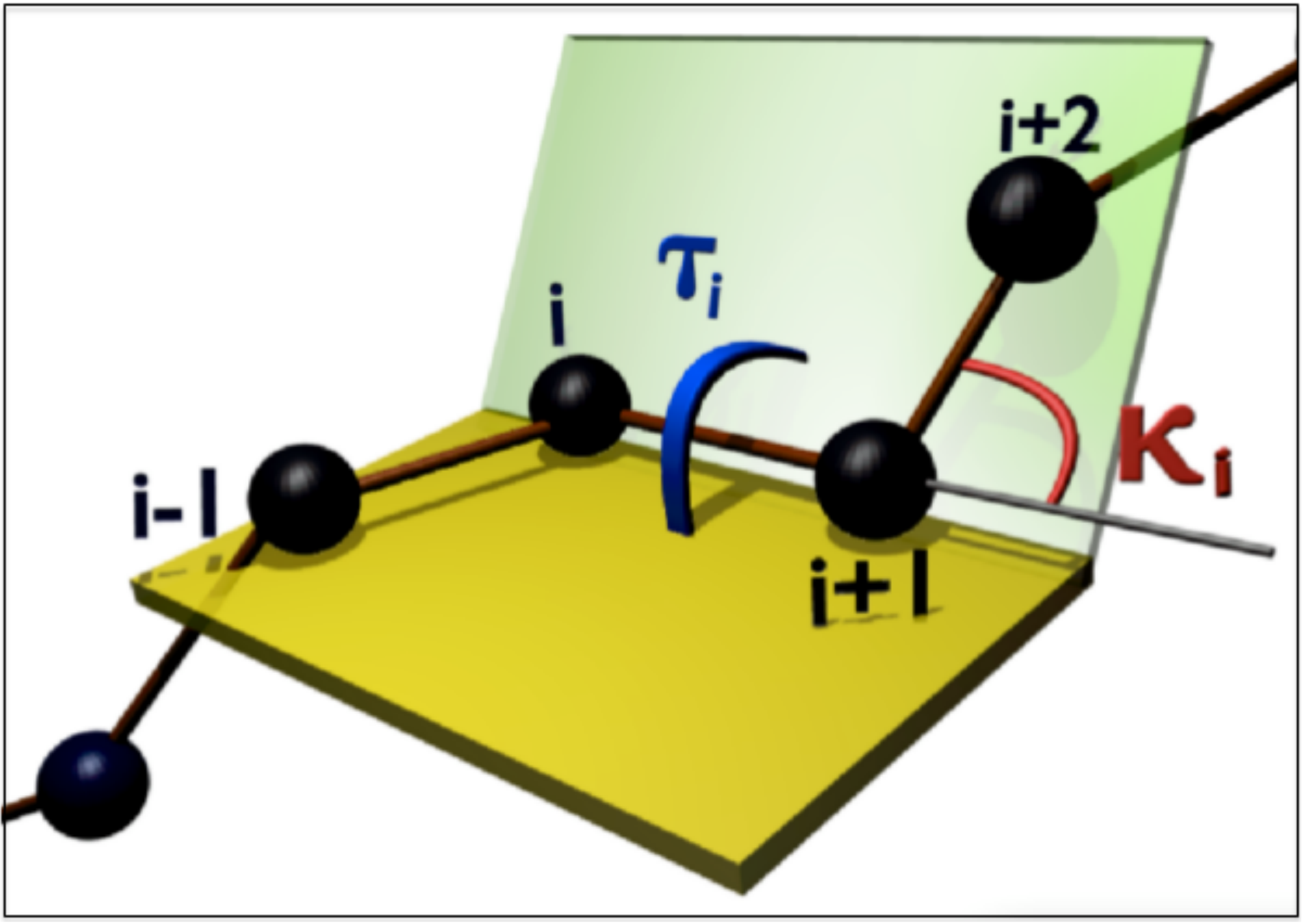}}
\end{center}
\vfill
\hfill{\large\sf Figure \ref{fig-2}}

\clearpage
\begin{center}
\resizebox{15.cm}{!}{\includegraphics[]{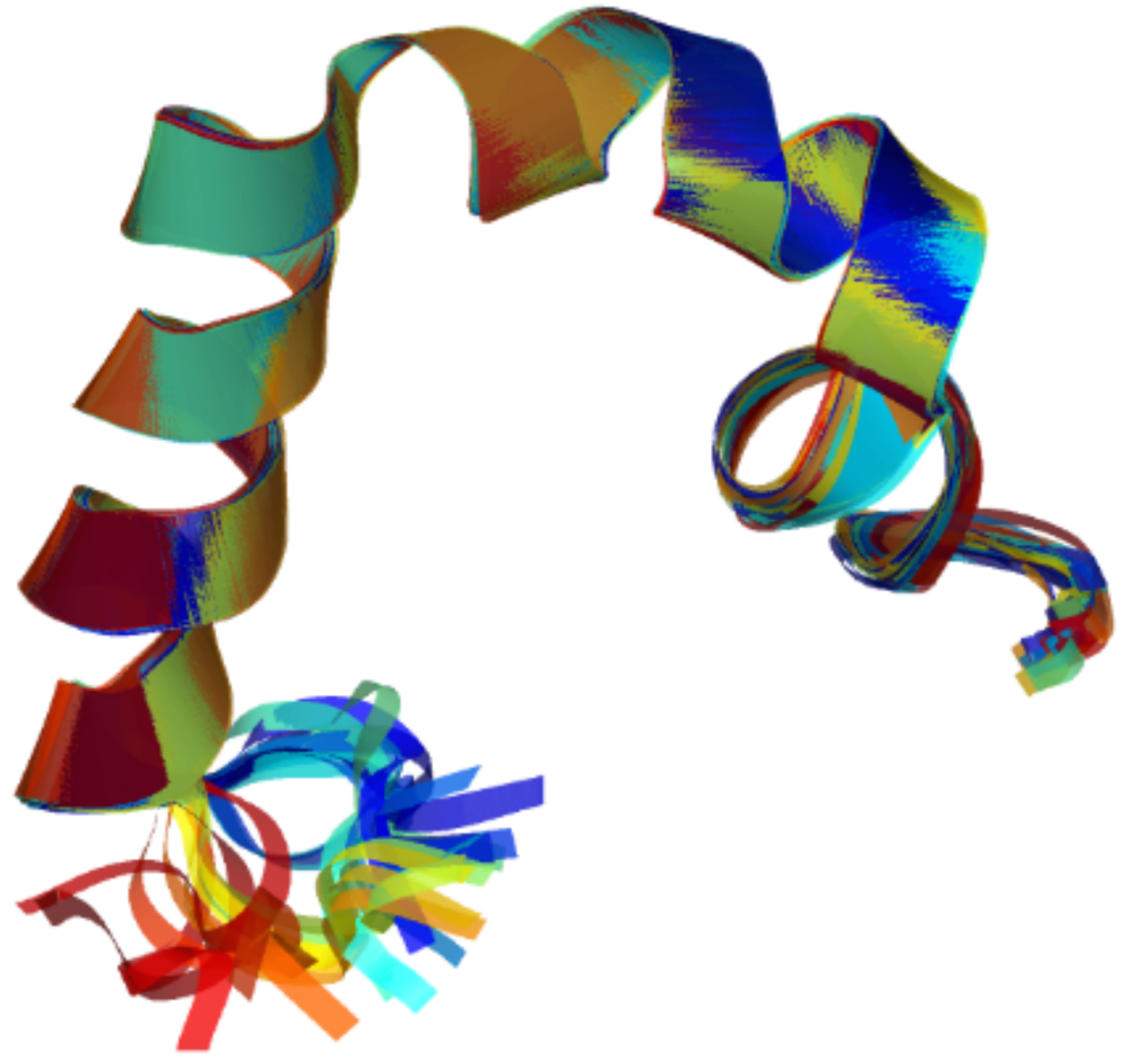}}
\end{center}
\vfill
\hfill{\large\sf Figure \ref{fig-3}}

\clearpage
\begin{center}
\resizebox{15.cm}{!}{\includegraphics[]{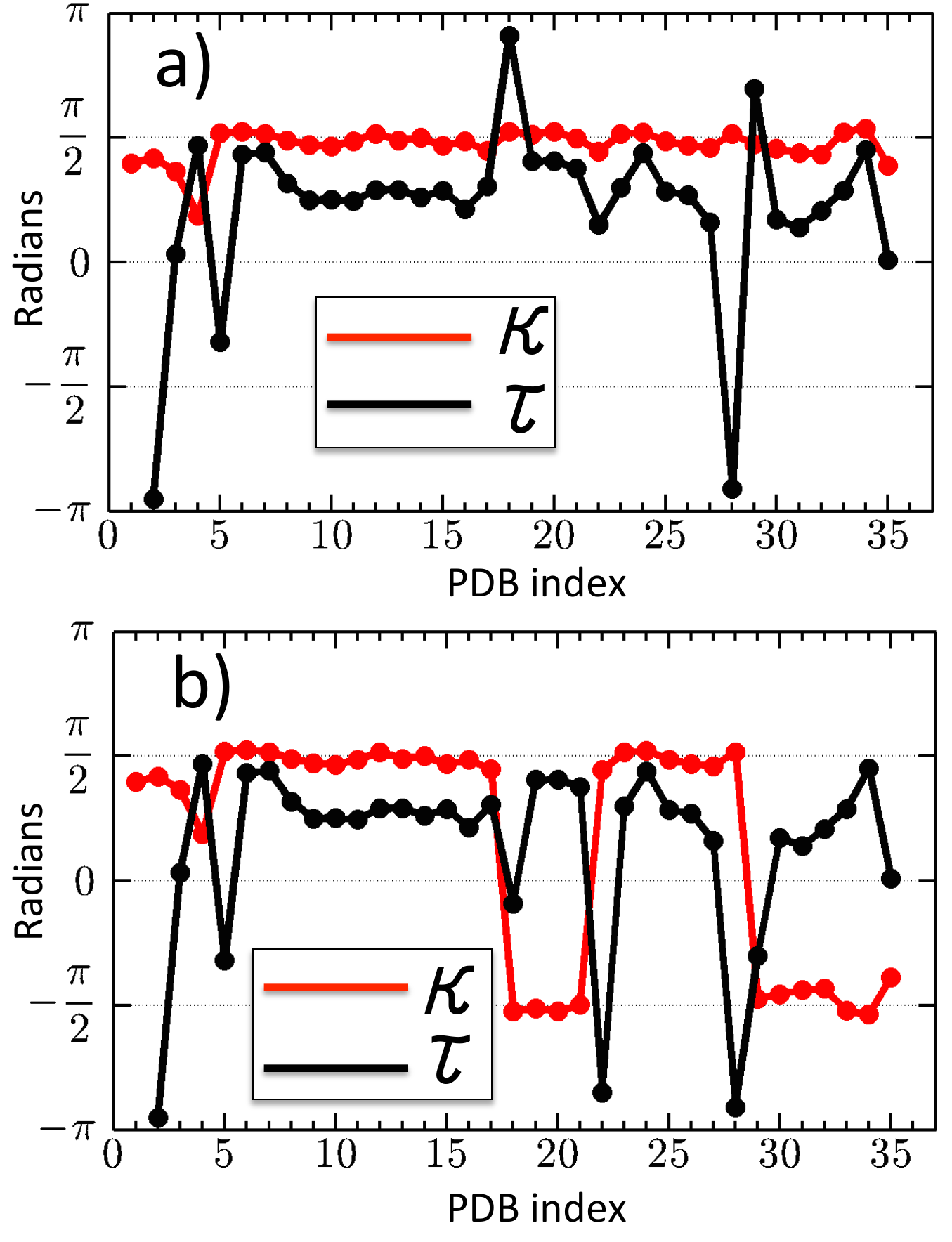}}
\end{center}
\vfill
\hfill{\large\sf Figure \ref{fig-4}}

\clearpage
\begin{center}
\resizebox{15.cm}{!}{\includegraphics[]{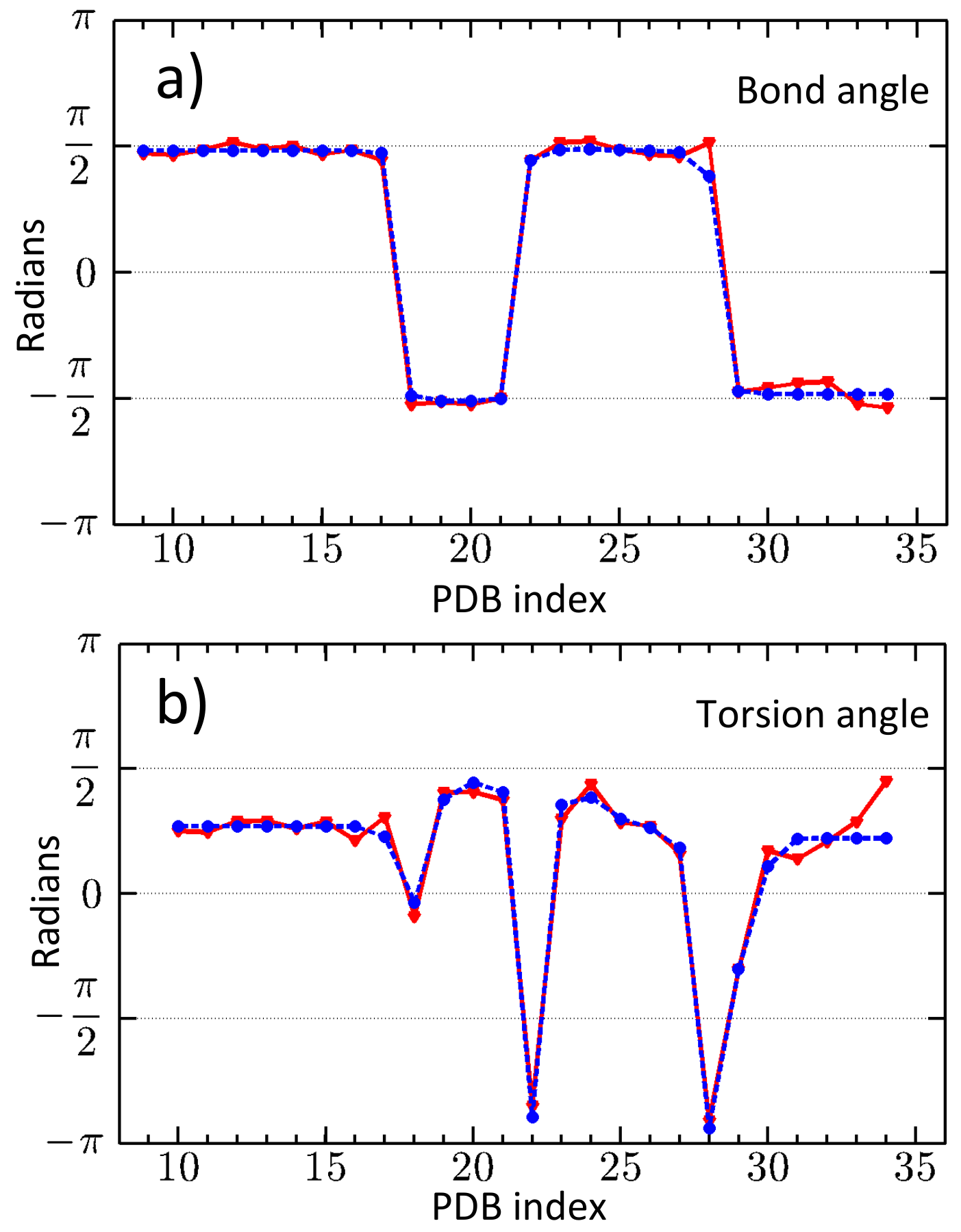}}
\end{center}
\vfill
\hfill{\large\sf Figure \ref{fig-5}}

\clearpage
\begin{center}
\resizebox{15.cm}{!}{\includegraphics[]{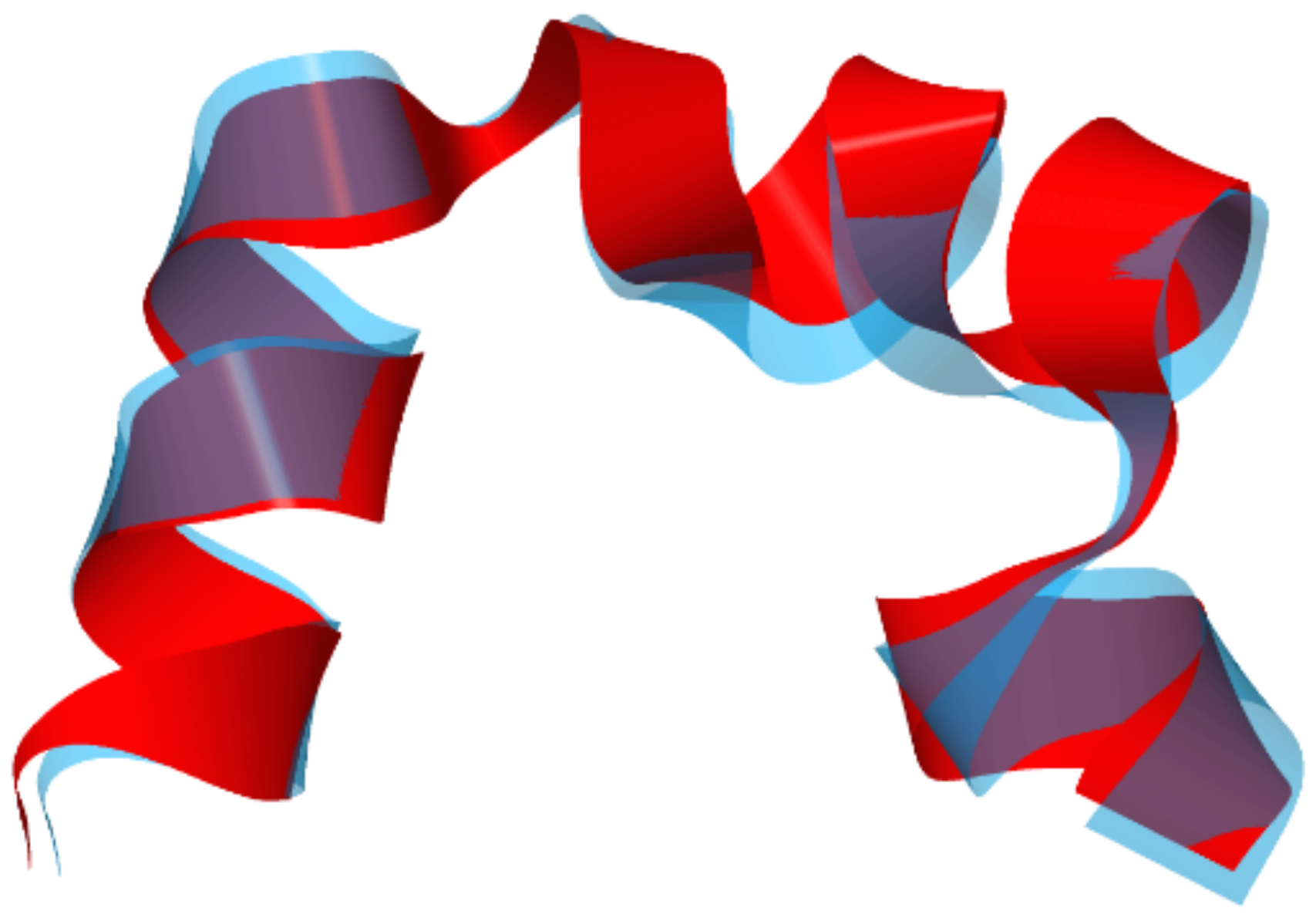}}
\end{center}
\vfill
\hfill{\large\sf Figure \ref{fig-6}}

\clearpage

\begin{center}
\resizebox{15.cm}{!}{\includegraphics[]{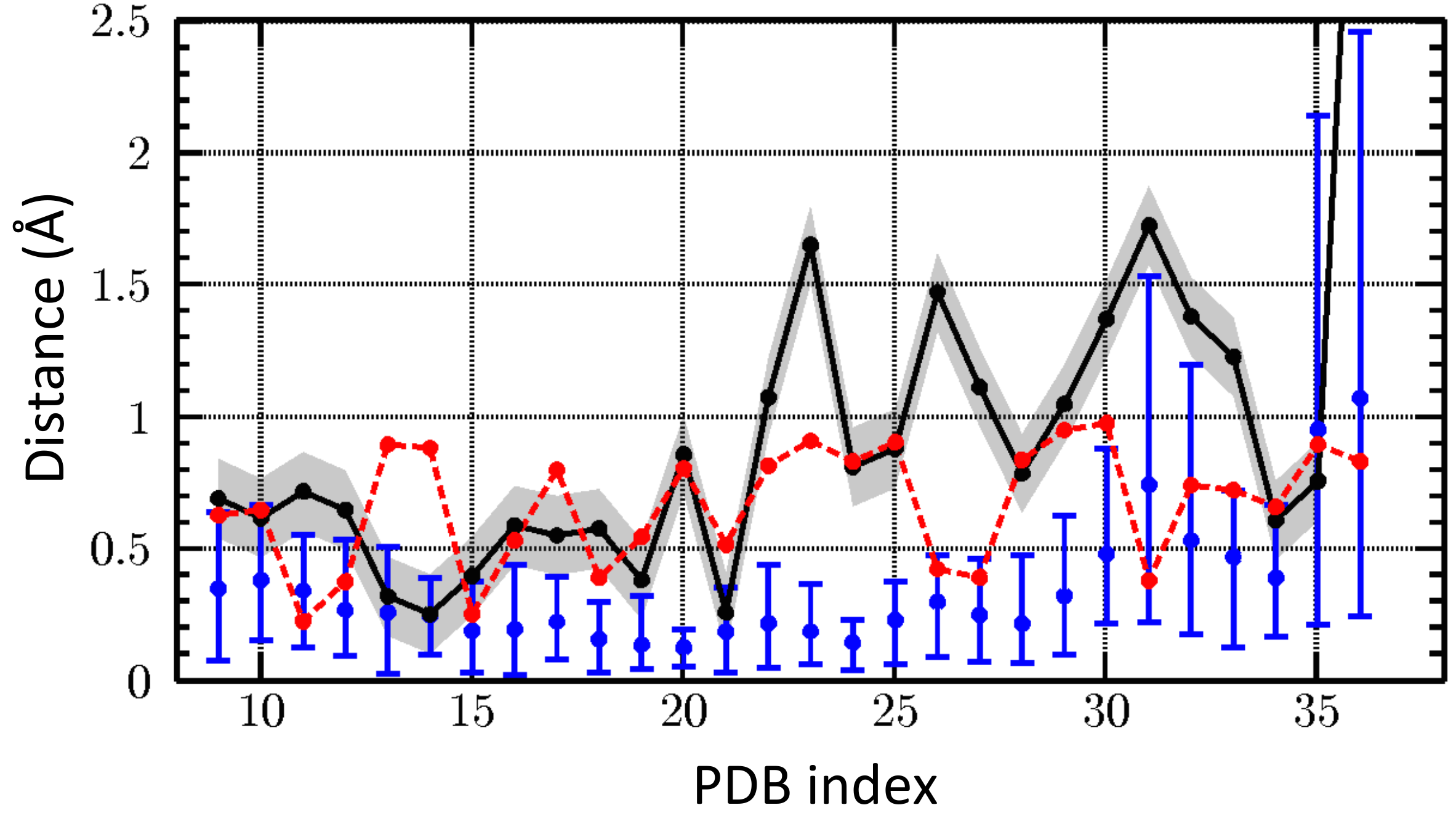}}
\end{center}
\vfill
\hfill{\large\sf Figure \ref{fig-7}}

\clearpage

\begin{center}
\resizebox{15.cm}{!}{\includegraphics[]{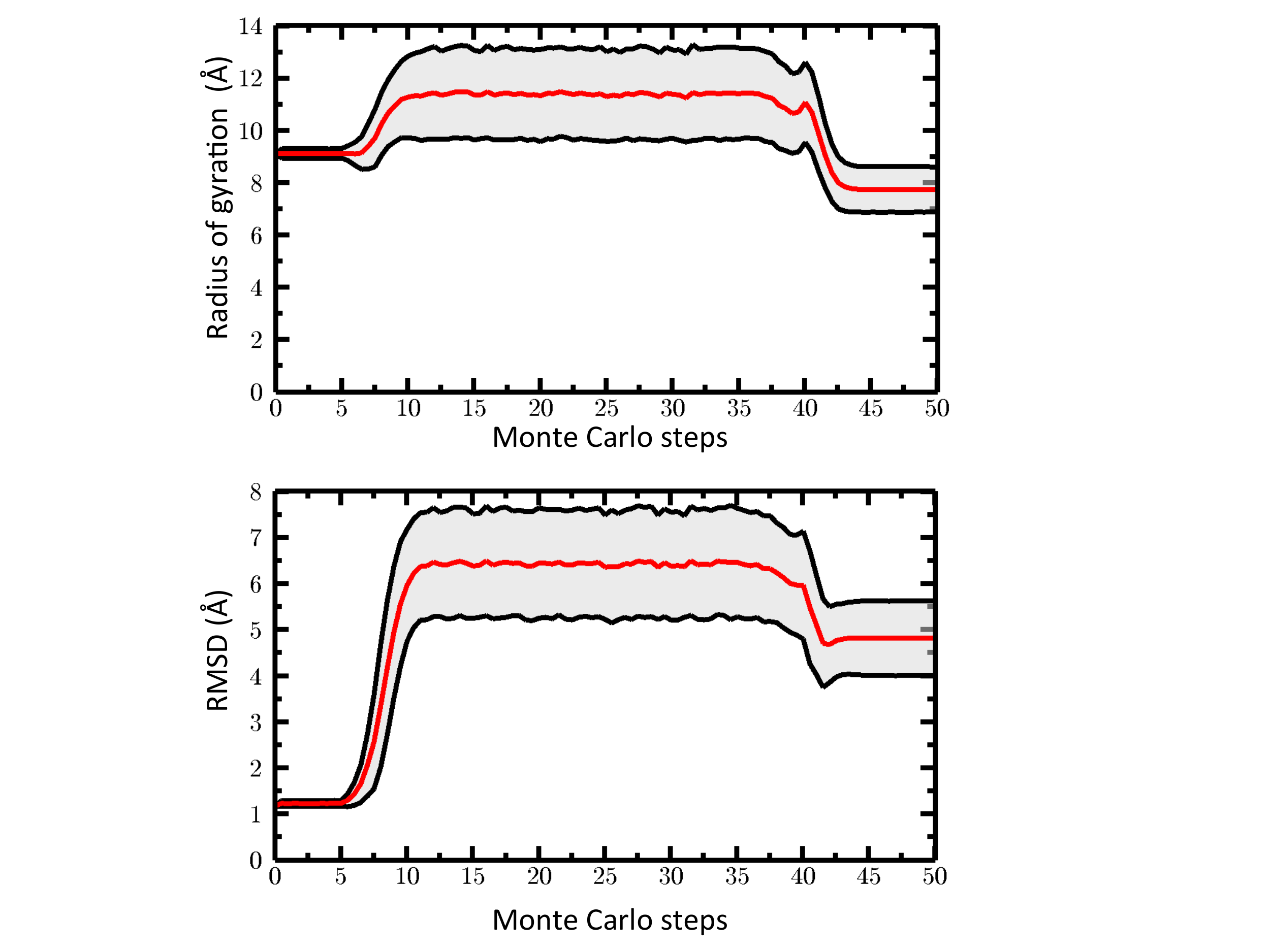}}
\end{center}
\vfill
\hfill{\large\sf Figure \ref{fig-8}}

\clearpage

\begin{center}
\resizebox{15.cm}{!}{\includegraphics[]{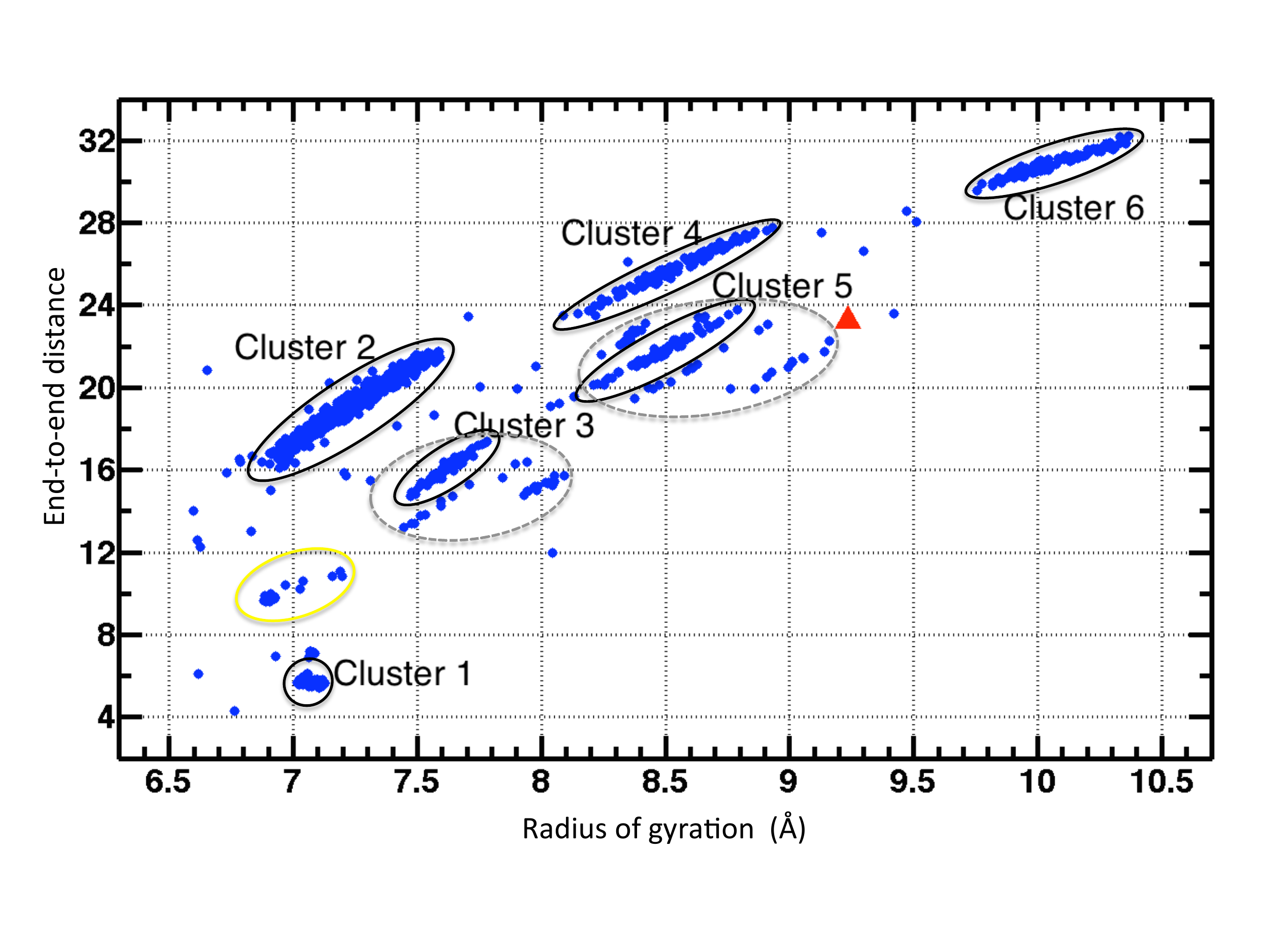}}
\end{center}
\vfill
\hfill{\large\sf Figure \ref{fig-9}}

\clearpage

\begin{center}
\resizebox{15.cm}{!}{\includegraphics[]{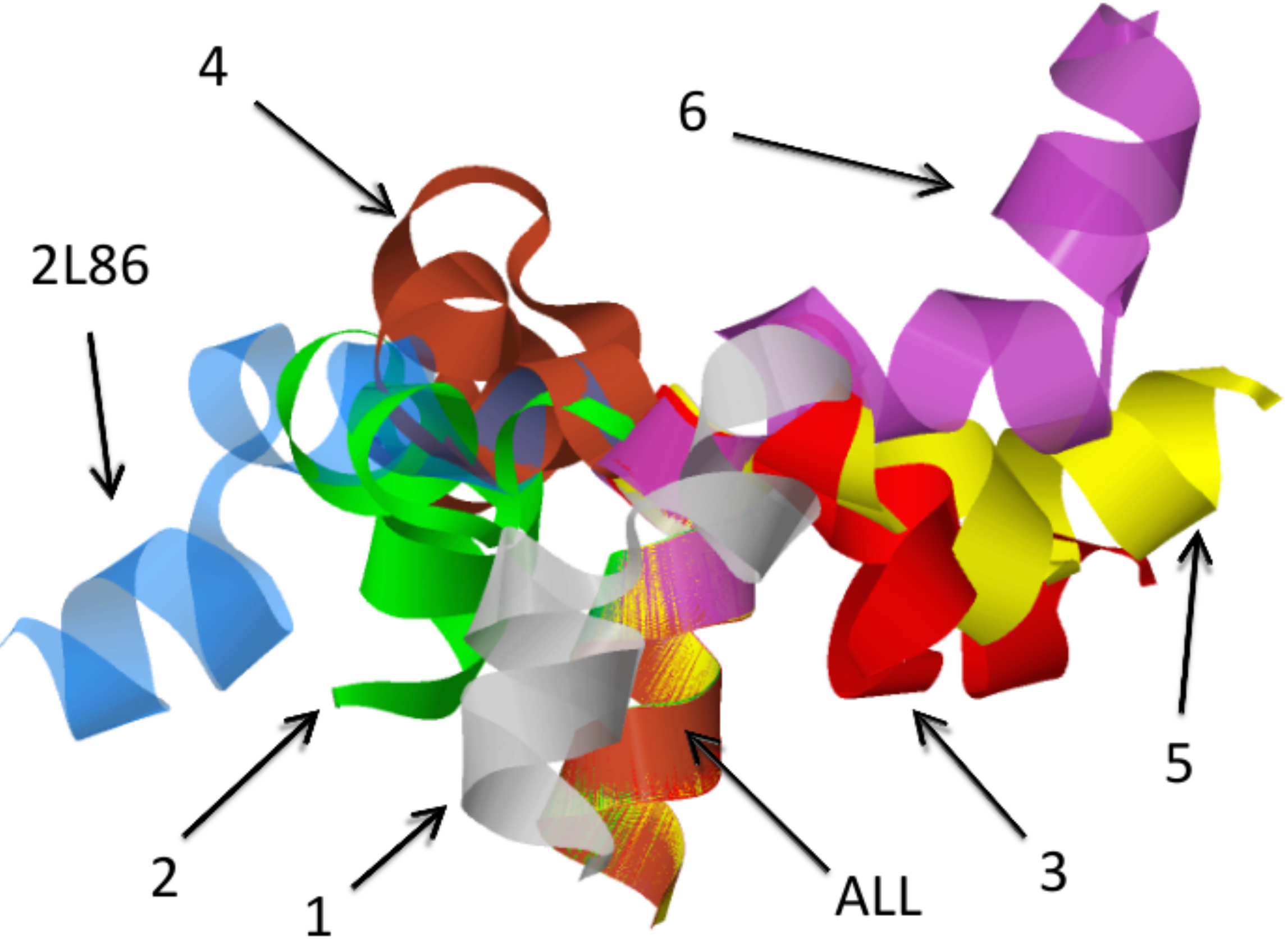}}
\end{center}
\vfill
\hfill{\large\sf Figure \ref{fig-10}}

\clearpage

\begin{center}
\resizebox{15.cm}{!}{\includegraphics[]{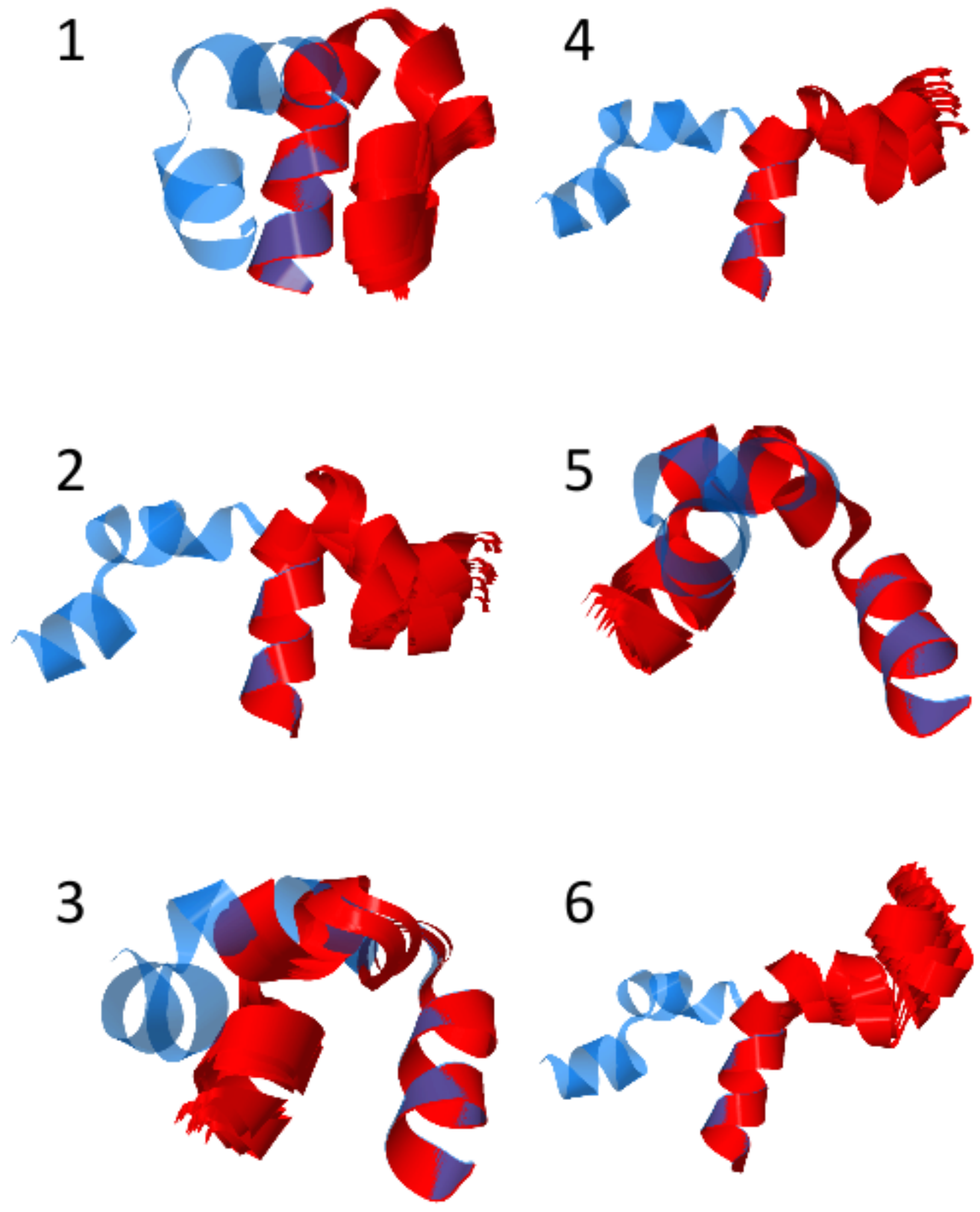}}
\end{center}
\vfill
\hfill{\large\sf Figure \ref{fig-11}}

\clearpage

\begin{center}
\resizebox{15.cm}{!}{\includegraphics[]{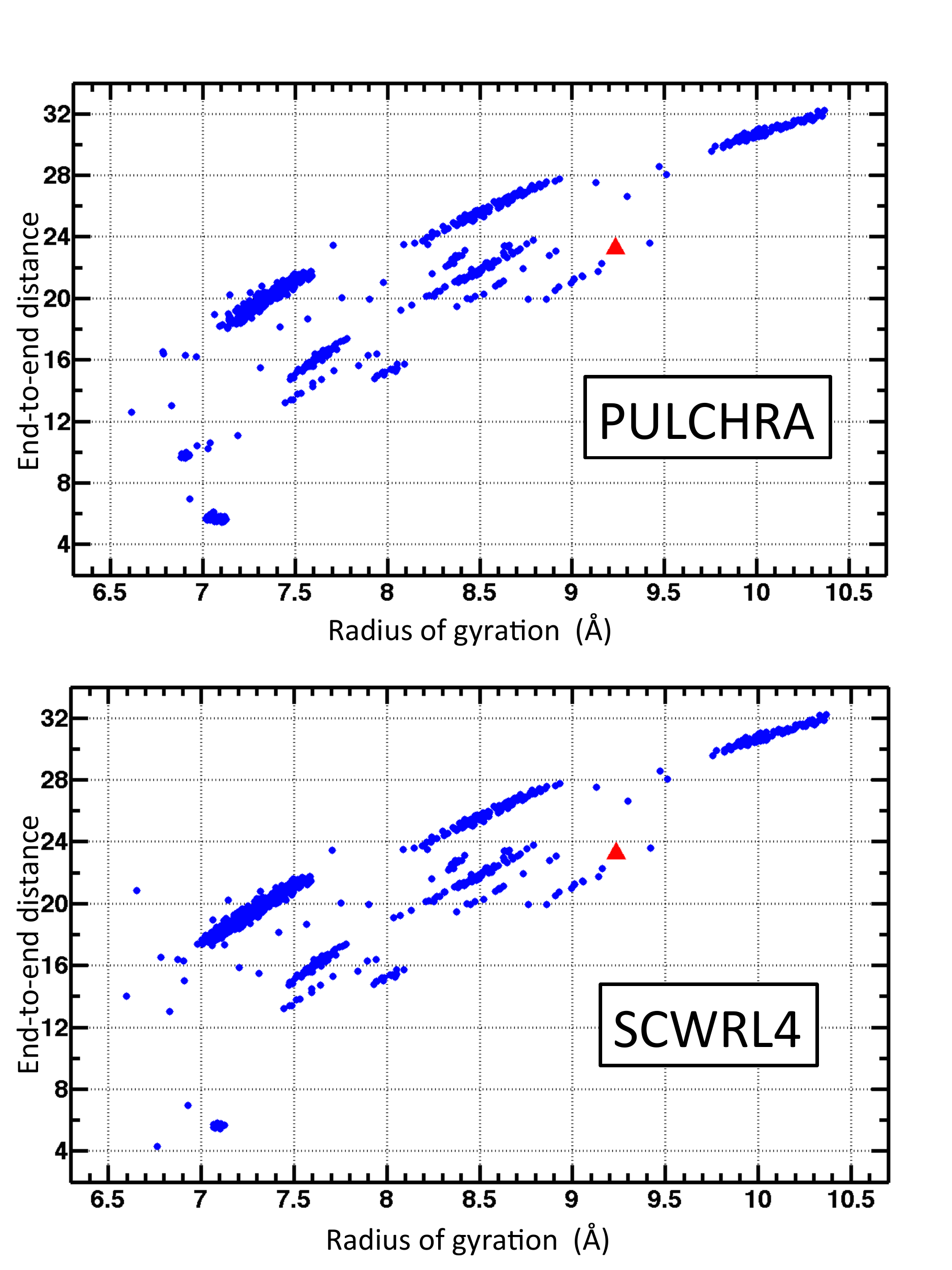}}
\end{center}
\vfill
\hfill{\large\sf Figure \ref{fig-12}}

\clearpage

\begin{center}
\resizebox{8.cm}{!}{\includegraphics[]{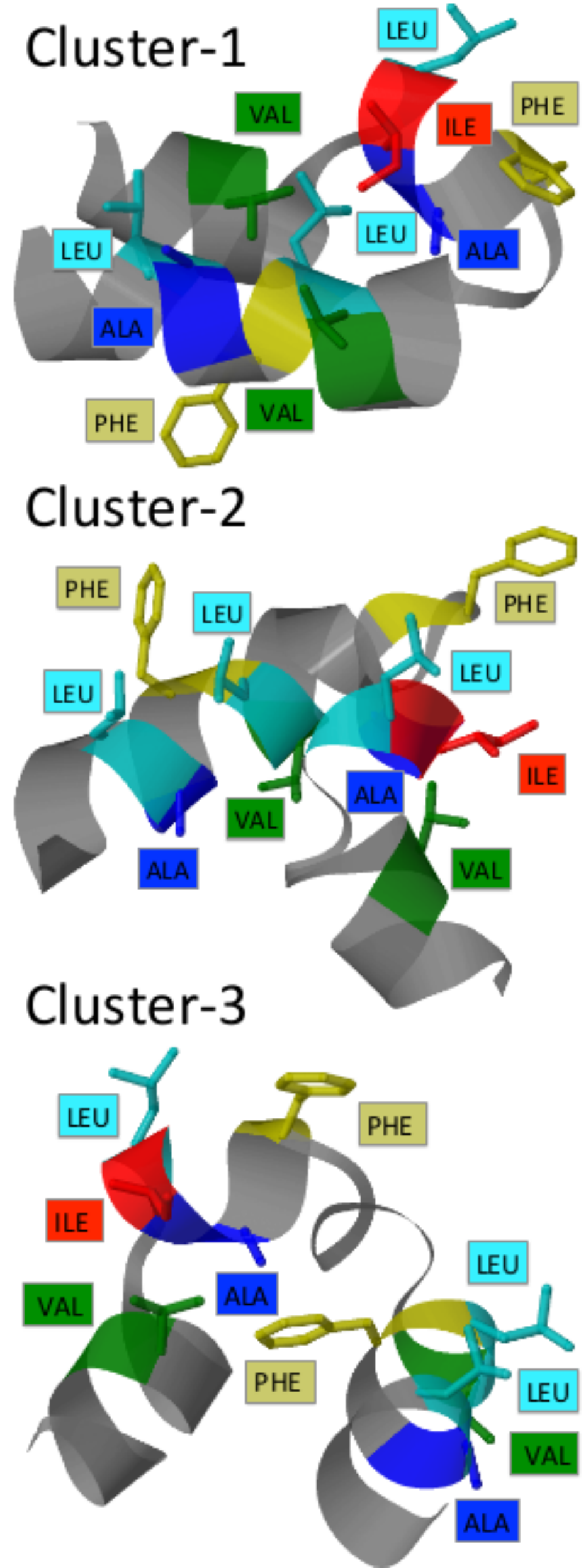}}
\end{center}
\vfill
\hfill{\large\sf Figure \ref{fig-13}}

\clearpage

\begin{center}
\resizebox{8.cm}{!}{\includegraphics[]{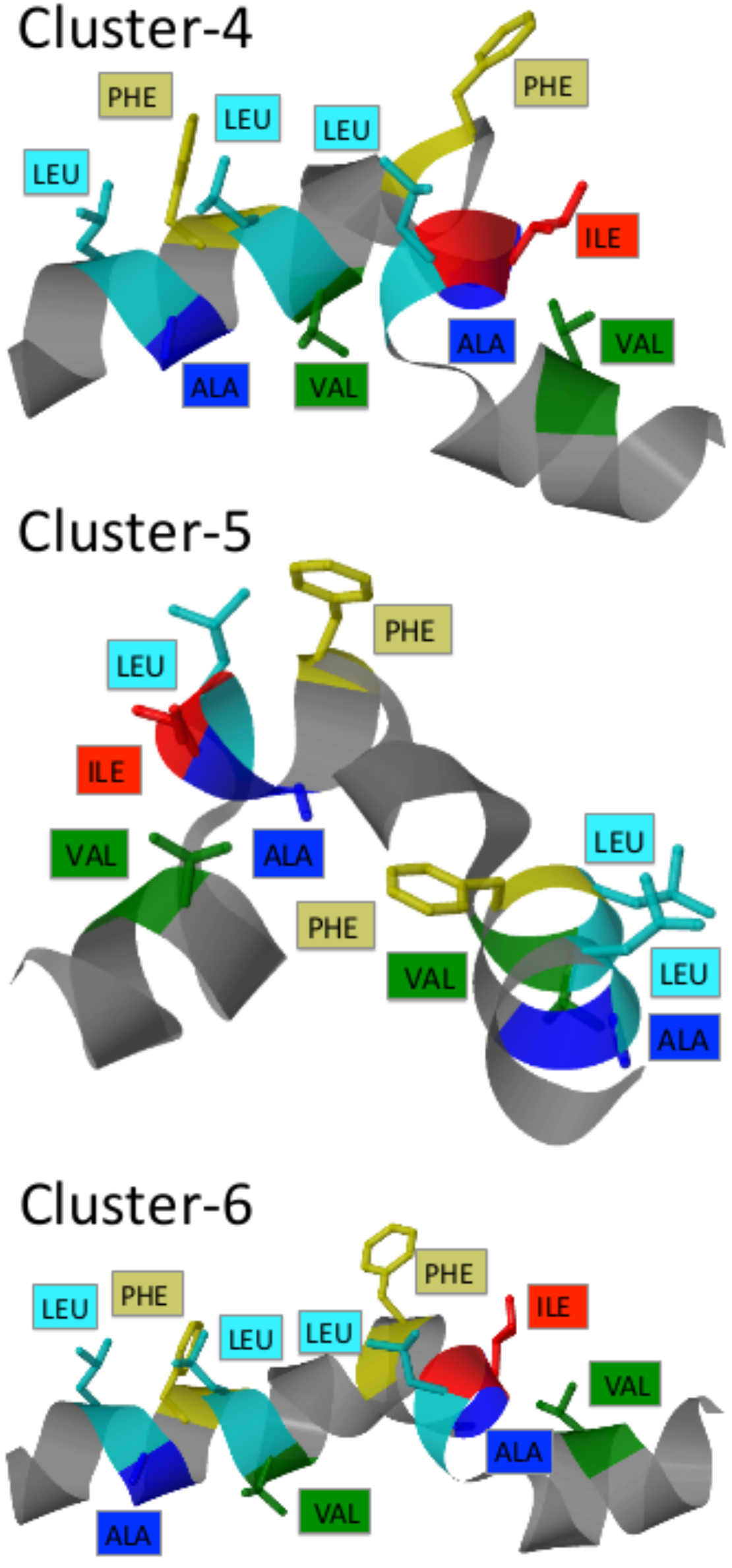}}
\end{center}
\vfill
\hfill{\large\sf Figure \ref{fig-14}}

\clearpage

\clearpage

\begin{table}[tbh]
\begin{center}
\caption{Parameter values for the three-kink configuration that describes 2L86; the soliton 1 cover the PDB segment T 9 -- N 21, the  soliton 2 covers the segment N 22 -- A 25 and the third soliton covers the segment I 26 -- T 36.  The value of  $a$ is fixed to
$a=-10^{-7}$.}
\vspace{10mm}
\begin{tabular}{|c|ccccccc|}
\hline 
soliton & $q_1$ & $q_2$ & $m_1$ & $m_2$ & $d/a$ & $c/a$ & $b/a$ \\
\hline
 1 & ~9.45452 ~&~ 4.45398 ~&~ 1.52110 ~&~ 1.60621 ~&~-8.1644$\cdot 10^{-2}$ ~&~ 
-1.4025$\cdot 10^{-3}$ ~&~ -2.5688~ \\
\hline
2 & ~2.92700 ~&~ 2.44082 & 1.66774 & 1.53420 & -4.8946$\cdot 10^{-1}$ &  -1.0672$\cdot 10^{-3}$ &- 19.4873~ \\
\hline
3 & ~1.11956 ~&~ 8.08649 & 1.52286 & 1.51493 & -3.5784$\cdot 10^{-2}$ & -5.9078$\cdot 10^{-3}$ & - 1.9085~
\\
\hline
\end{tabular}
\end{center}
\label{Table I}
\end{table}

  \end{document}